\documentstyle[12pt]{article}

\newcommand{\bb}{\begin{equation}}
\newcommand{\ee}{\end{equation}}

\begin{document}
\rightline{ULB-TH 11/94}

{}~\\
{}~\\
\begin{center}
{\huge {\bf A note on the path integral for systems with
primary and secondary second class constraints}}\\
\vspace{2.5cm}
{\large Marc Henneaux$^{a,b}$} and
\vspace{1cm}
{\large Andrei Slavnov$^{a,c}$}\\\vspace{1.5cm}
{}~$^a$ {\em Facult\'e des Sciences, Universit\'e
Libre de Bruxelles, Campus Plaine C.P. 231, B-1050
Bruxelles, Belgium.} \\
{}~$^b${\em Centro de Estudios Cient\'{\i}ficos de Santiago,
Casilla 16443, Santiago 9, Chile} \\
{}~$^c${\em Steclov Mathematical Institute, Vavilov 42,
GSP-1, 117966, Moscow} \\

\vspace{2cm}
\end{center}

\begin{abstract}

\noindent
It is shown that the phase space path integral for a
system with arbitrary second class constraints (primary,
secondary ...) can be rewritten as a configuration space path
integral of the exponent of the Lagrangian action with some local
measure.
\end{abstract}

\vfill
\break

\newpage

The path integral for a system $(q^i, p_i)$ with arbitrary second
class constraints $\chi_a = 0$ (primary, secondary ...) and
Hamiltonian $H$ is given by
\bb
Z = \int {\cal D}q {\cal D}p \prod_t |det[\chi_a, \chi_b]|^{1/2}
\, \delta (\chi_a) \, \exp{i\int (p_iJ\dot q^i - H) dt}
\ee
(see \cite{Faddeev,Senj} and \cite{HT} for a recent review).  The purpose
of this note is to show that one can rewrite (1) as the Lagrangian
path integral
\bb
Z = \int {\cal D}q \rho_L \exp{iJ\int L dt},
\ee
where $L$ is the original Lagrangian and where $\rho_L$
is a {\bf local} measure.  Although this result is probably
known for some particular cases (free massive
vector field), it has not been proved, to our knowledge, in
full generality.  This is done in this letter.

Our paper is divided in three parts.  First, we explain the difficulty
in going from (1) to (2).  Second, we establish (2).  And finally, we
comment on the usefulness and applications of the results.

\vspace{0.7cm}
\noindent
{\bf Difficulty :}  The difficulty in rewriting the Hamiltonian
path integral (1) as the Lagrangian path integral (2) stems from the
secondary, tertiary ... constraints.  Indeed, the path integral (1)
is equal to
\bb
Z = \int {\cal D}q {\cal D}p {\cal D}\lambda \prod_t
|det[\chi_a, \chi_b]|^{1/2} \exp{i \int (p_i \dot q^i - H -
\lambda^a \chi_a ) dt}J\, ,
\ee
where the $\delta$-functions of the constraints have
been exponentiated by means of Lagrange multipliers.
Now, if there were only primary constraints, one could solve the
equations $\delta S_H / \delta p_i = 0$ and
$\delta S_H / \delta \lambda^a = 0$ for $p_i$ and $\lambda^a$, which
appear thus as auxiliary fields.  The value of $S_H$ at the extremum
for $p_i$ and $\lambda^a$ is precisely the Lagrangian action $S_L$.
Therefore, if one evaluates (3) by stationary phase, one gets
precisely (2) where the local measure is the product of
$\prod_t |det[\chi_a, \chi_b]|^{1/2}$ evaluated at the extremum
 times the contributions coming from the integration over
$p_i$ and $\lambda^a$.  The same argument does not apply as such if there
are secondary, tertiary, ... constraints since it is then in general
impossible to solve the equations $\delta S_H / \delta p_i = 0$ and
$\delta S_H / \delta \lambda^a = 0$ for $p_i$ and $\lambda^a$.
Therefore, it is not clear that one can rewrite (1) as a path integral over
trajectories in configuration space of the exponential of $i$ times
the Lagrangian action.  Nevertherless, we shall show that this
statement is correct by making an appropriate change of variables.

\vspace{0.7cm}
\noindent
{\bf Solution :}  Let us first consider for definiteness the case of
one primary second class constraint $\chi_1 = 0$ and one secondary
second class constraint $\chi_2 = 0$.  We can always choose the
secondary constraint so that
\bb
[H, \chi_1] = \chi_2
\ee
and, of course,
\bb
[\chi_1, \chi_2] \not= 0.
\ee
The Hamiltonian path integral (3) is
\bb
Z = \int {\cal D}q {\cal D}p {\cal D}\lambda {\cal D}\mu \prod_t
|det[\chi_a, \chi_b]|^{1/2} \exp{i \int (p_i \dot q^i - H -
\lambda \chi_1 - \mu \chi_2 ) dt}
\ee
We now make the canonical change of variables $(q^i, p_i)$ $\rightarrow$
$(q^{\prime i}, p^\prime_i)$ generated by $\mu \chi_1$,
\bb
q^i \rightarrow q^{\prime i} = \exp{([\mu \chi_1)}, q^i]
\;\;  \equiv q^i + [\mu \chi_1, q^i] + \frac{1}{2} [\mu \chi_1,
[\mu \chi_1, q^i]] + \dots \, ,
\ee
\bb
p_i \rightarrow p^\prime_i = \exp{([\mu \chi_1)}, p_i] \, .
\ee
Under this canonical transformation, the measure, the kinetic term
$p \dot q$ in the action, and $\chi_1$ are invariant.  The Hamiltonian
becomes
\bb
H \rightarrow H^\prime = \exp{([\mu \chi_1)}, H] = H - \mu \chi_2 -
\frac{1}{2} \mu^2 [\chi_1, \chi_2] + O(\mu^3).
\ee
The secondary constraint transforms also as
\bb
\chi_2 \rightarrow \chi^\prime_2 = \exp{([\mu \chi_1)}, \chi_2] =
\chi_2 + \mu [\chi_1, \chi_2] + O(\mu^3) \, .
\ee
Thus, in terms of the new variables, the path integral reads
\begin{eqnarray}
Z = \int {\cal D}q {\cal D}p {\cal D}\lambda {\cal D}\mu \prod_t
|det[\chi_a, \chi_b]|^{1/2} \; \; \; \; \; \; \; \; \; \; \; \; \;
\; \; \; \; \; \; \; \; \; \; \; \; \; \; \; \; \; \; \; \; \;
\; \; \; \; \; \;   \nonumber \\
\; \; \; \; \; \; \; \; \; \; \; \;
 \times \exp{i \int (p_i \dot q^i - H -
\lambda \chi_1 - \frac{1}{2} \mu^2 [\chi_1, \chi_2] + O(\mu^3) ) dt} \, .
\end{eqnarray}
One sees that the term linear in $\mu$ has disappeared from the
action.  The integration over $\mu$ can be done by stationary
phase since the coefficient of $\mu^2$ is different from zero.  This yields
\bb
Z = \int {\cal D}q {\cal D}p {\cal D}\lambda  \rho
\exp{i \int (p_i \dot q^i - H -
\lambda \chi_1 ) dt} \, .
\ee
Here, $\rho$ is some local measure.  It is given explicitly by
\bb
\rho = \prod_t |det[\chi_1, \chi_2]|^{1/2}
\ee
when the series in $\mu$ in (11) terminates at the second order.  In
general, there are additional local contributions to the measure
(13) which are calculable order by order in the stationary phase
expansion.  In the path integral (12), only the primary constraint
appears.  We can thus repeat the derivation
given above and integrate over
$p_i$ and $\lambda$ to reach
the desired Lagrangian form.

Obviously, the procedure directly generalizes if there are more than one
pair of primary and secondary second class constraints.
One can also generalize the derivation to the case when constraints of
higher generation (tertiary, quaternary ...) are present.  This is done
as above, by making a change of variables that eliminates the terms linear
in the Lagrange multipliers associated with the secondary, tertiary ...
constraints.  We shall assume that there is an equal number of constraints
at each generation.  One can show \cite{Lusanna} that the general
case can be reduced to this one, in the sense that there may be
various chains of generations of
different lengths, but that these commute in the Poisson bracket.
The Hamiltonian path integral reads then
\begin{eqnarray}
Z = \int {\cal D}q {\cal D}p {\cal D}\lambda^{(1)} {\cal D}\lambda^{(2)}
\dots {\cal D}\lambda^{(L)} \prod_t
|det[\chi_a, \chi_b]|^{1/2} \; \; \; \; \; \; \; \; \; \; \; \; \;
\; \; \; \; \; \; \; \; \; \;
\; \; \;   \nonumber \\
 \times \exp{i \int (p_i \dot q^i - H -
\lambda^{(1) \alpha} \chi^{(1)}_\alpha -
\lambda^{(2) \alpha} \chi^{(2)}_\alpha  - \dots -
\lambda^{(L) \alpha} \chi^{(L)}_\alpha ) dt} \, ,
\end{eqnarray}
where $\chi^{(1)}_\alpha$ are the primary constraints, $\chi^{(2)}_\alpha$
are the secondary constraints, $\chi^{(3)}_\alpha$ are the
tertiary constraints ..., chosen so that
\bb
\chi^{(i)}_\alpha = [ H, \chi^{(i-1)}_\alpha] \; , i=2,3,\dots ,L
\ee
where $L \geq 2$ is the number of generations.

The transformation that eliminates the linear terms is
a straightforward generalization of (7), (8) and reads
explicitly
\bb
F(q,p) \rightarrow F^\prime = \exp{([\sum_{k=1}^{L-1} \lambda^{(k+1) \alpha}
 \chi^{(k)}_\alpha)}, F] \, .
\ee
The quadratic term generated by the transformation is
\bb
\frac{1}{2} [ \sum_{k=1}^{L-1} \lambda^{(k+1) \alpha}
 \chi^{(k)}_\alpha, \sum_{k=2}^{L} \lambda^{(k) \alpha}
 \chi^{(k)}_\alpha ]
\ee
and is easily verified to be non degenerate using the canonical
representation of the brackets given in \cite{Lusanna}.  One can
thus evaluate the integral over $\lambda^{(2) \alpha}$,
$\lambda^{(3) \alpha}$, ..., $\lambda^{(L) \alpha}$ by stationary
phase, which results in a mere modification in the local measure
in the path integral.  Then, one can integrate over the momenta
$p_i$ and the Lagrange multipliers $\lambda^{(1) \alpha}$ of
the primary constraints to get the Lagrangian path integral (2).

\vspace{0.7cm}
\noindent
{\bf Comments :}J The advantages of the Lagrangian path integral are
twofold.  First, we avoid having to solve explicitly the higher
generation constraints.  Second, in the case of a relativistic theory,
the exponent in the path integral is manifestly invariant, while
secondary or higher generation constraints usually spoil manifest
Lorentz invariance.

As we have already discussed, there is in (2) a
non trivial local measure which can formally be represented as
\bb
\rho_L = \exp{i \int dt (\delta (0) M_1 + \delta^2(0) M_2 + \dots ) } \, .
\ee
Of course, this expression is not well defined.  To make sense
out of it, one should specify some regularization method.
If one can use dimensional regularization, all these terms vanish and the
Lagrangian local measure $\rho_L$ is equal to unity.  In that case,
we arrive at a very simple and complete expression for the
Lagrangian path integral,
\bb
Z = \int {\cal D}q \exp{i \int dt L(q,J\dot q)} \, ,
\ee
which does not require the knowledge of the Dirac bracket.  If, however,
dimensional regularization - or some other regularization that
makes $\delta(0)$ equal to zero - is not available, then one must define
the meaning of (18) for the model under consideration.

The form of expression (18) for the local measure should not be taken
too literally.  In particular, it may not be covariant.  However, it
does not really mean the breaking of Lorentz invariance because the
calculation of the path integral may produce analogous terms
compensating the non invariance of the measure.  This phenomenon
may actually occur already for unconstrained systems.  For instance,
let us consider the Lagrangian
\bb
{\cal L} = \frac{1}{2} \partial_\mu \phi \partial^\mu \phi
+ \frac{1}{2} g (\partial_\mu \phi A^\mu )^2
\ee
where $A^\mu$ is treated as an external field.  The canonical quantization
of this system leads to the well-known path integral
\bb
Z = \int {\cal D}\phi \prod_t |det(1 + g (A^0)^2)|^{\frac{1}{2}} \,
\exp{i\int {\cal L} dx}\, .
\ee
If taken at face value, this expression is not a Lorentz-invariant functional
of $A_\mu$.  However, careful calculation of the path integral shows
that the correlation functions of $\partial_\mu \phi (x) \partial_\nu
\phi (y)$ contains non covariant contributions proportional to
$\delta_{\mu 0} \delta_{\nu 0} \delta (x-y)$ which compensate the
non covariant contribution from the measure.  This is a phenomenon
characteristic of
theories with derivative couplings.

Another simple model which exhibits the same features
and which involves second class constraints  is described by the
Lagrangian
\bb
{\cal L} = g \partial_\mu \phi A^\mu (A^\nu A_\nu) + \frac{1}{2} \phi^2
+ \frac{1}{2}J\partial_\mu A_\nu \partial ^\mu A^\nu \, .
\ee
This model has both primary and secondary second class constraints.
Application of the previous method leads to the Lagrangian path
integral
\bb
Z = \int {\cal D}\phi {\cal D}A \prod_t |det((A_\nu A^\nu)^2 +
 8 A_0^2 A_\nu A^\nu)|^{\frac{1}{2}} \,
\exp{i\int {\cal L} dx}\, .
\ee
As in the previous example, the local measure is not Lorentz invariant
but this non invarianve is again compensated by singular contributions
coming from the Lagrangian.

We close this letter with two additional remarks.  First, we
considered here only the path integral.  A complete definition
of the quantum theory necessitates of course the knowledge of the
physical states, for which the full constraint spectrum
(including the secondary, tertiary ...constraints)
and Dirac bracket machinery are relevant.  Second, one may easily
include first class constraints.  This is because first class
constraints commute with second class ones.  One finds
again complete agreement between the Hamiltonian path
integral and the Lagrangian path integral obtained by standard
Lagrangian tools (e.g. antifield formalism), provided one
includes in the latter an appropriate local measure.

\vspace{2cm}
\noindent
{\large {\bf Acknowledgements}}

This work has been supported in
part by a research grant from F.N.R.S., by research contracts
with the Commission of the European Community, by ISF-grant
MNB000 and by the Russian Fund for Fundamental studies
under grant number 94-01-003000a.

\end{document}